\documentclass[aps,prb,twocolumn,floats]{revtex4}
\usepackage{epsfig}
\usepackage{graphicx}
\usepackage{amsmath}
\usepackage{multirow}

\begin{document}
\topmargin -4mm

\title{Electronic Structures of Porous Nanocarbons}
\author{Artem Baskin and Petr Kr\'al$^*$}

\affiliation{Department of Chemistry, University of Illinois at Chicago,
Chicago, IL 60607, USA}

\vspace{4mm}


\vspace{10mm}

\begin{abstract}
We use large scale {\it ab-initio} calculations to describe electronic
structures of graphene, graphene nanoribbons, and carbon nanotubes periodically
perforated with nanopores. We disclose common features of these systems
and develop a unified picture that permits 
us to analytically predict and systematically characterize 
metal-semiconductor transitions in nanocarbons with superlattices of
nanopores of different sizes and types. These novel materials with highly
tunable band structures have numerous potential applications in 
electronics, light detection, and molecular sensing.
\end{abstract}

\maketitle
Graphene \cite{Novoselov} has unique and highly tunable parameters, which 
can be exploited in novel hybrid materials and devices with numerous 
applications.  It can be modified by doping \cite{Palacios,Balog}, chemical 
functionalization \cite{Cervantes}, and geometrical restrictions, such as 
cutting and introduction of defects and pores \cite{Furst,Vanevic,Pedersen}. 
Recently, graphene perforated with nanopores was used as a selective sieve 
for hydrated ions \cite{Sint}, gases \cite{Jiang,Blankenburg}, and DNA 
\cite{Postma,Garaj}. 

In this work, we use large scale {\it ab-initio} calculations to describe
electronic structures of nanocarbons perforated with superlattices of 
nanopores.  We search for common principles allowing us to characterize the
electronic structures of {\it porous nanocarbons} (PNC), such as porous 
graphene (PG), porous graphene nanoribbons (PGNR), and porous carbon 
nanotubes (PCNT). Although, partial results for the electronic structures 
of PG \cite{Furst,Pedersen,Vanevic} and other PNCs have been obtained 
\cite{Gao,Fujita,Shima,Roche}, general principles that would unify their 
electronic structures are missing. 

\begin{figure}[t]
\hspace*{-2mm}
\hbox{\epsfxsize=87mm \epsffile{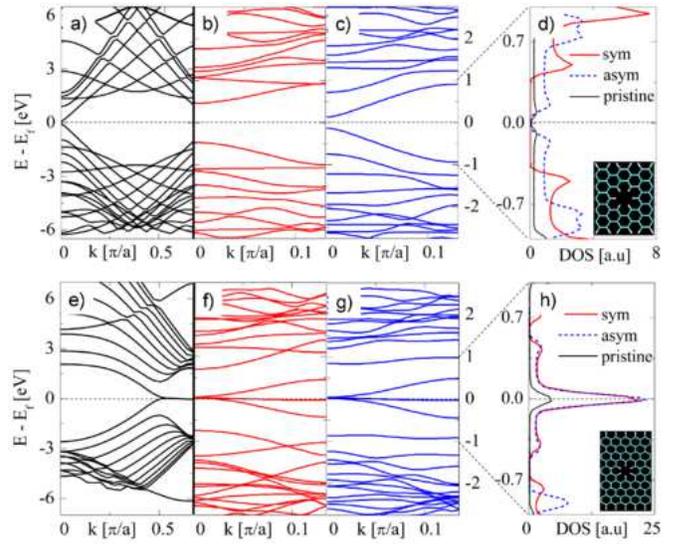}}
\vspace*{-6mm}
\caption{
\small Band structure of: a) pristine 11-AGNR, b) 11-AGNR with centered
SP, c) 11-AGNR with shifted SP, e) pristine 10-ZGNR, f) 10-ZGNR with 
centered SP, g) 10-ZGNR with shifted SP. The energy scales
for (b, c, f, g) cases are the same. Density of states: d) centered and 
hifted SP in 11-AGNR, h) centered and shifted SP in 10-ZGNR. Insets: 
Unit cells for 11-AGNR and 10-ZGNR with the standard pore}
\label{azpore}
\vspace*{-6mm}
\end{figure}

\vspace{2mm}
{\bf \Large Results}

The studied PNCs are perforated with pores of different shapes, 
sizes, and locations. Most of the results are obtained for arrays of 
honeycomb-shaped pores, called the ``standard pore" (SP) 
\cite{Blankenburg,Bieri}, where in each 
pore six C atoms (one benzene ring) are excluded from the studied nanocarbon
and the dangling bonds are terminated by hydrogens. All the PNCs have their 
edges and pores passivated by hydrogen atoms. The details of our calculations
are described in {\it Methods}.

{\it Porous graphene nanoribbons:\ }

We begin our study by examining the band structures of porous armchair (AGNR) 
and zigzag (ZGNR) graphene nanoribbons. All GNRs are classified by the number 
of carbon dimers, $N$, that form the ribbon ($N$-GNR) \cite{Roche}. First, we 
study the porous 11-AGNR and 10-ZGNR (both metallic when pristine), 
and elementary cells are shown in Figs.~\ref{azpore} (d, h). 
In Figs.~\ref{azpore} (a, b), we can see that the introduction of a periodic 
array of SPs in the center of 11-AGNR causes a significant band-gap opening 
(0.15 eV $\to$ 0.92 eV).
On the other hand, both pristine and porous (same pores) 10-ZGNR 
have no energy band gap, as seen in Figs.~\ref{azpore} (e, f). When the SPs
are displaced by one honeycomb cell towards the GNR-edge, the band gap 
in 11-AGNR shrinks $\approx$ 3.75 times, while 10-ZGNR remains metallic, 
as shown in Figs.~\ref{azpore} (c, g). We also checked that the PGNR band 
structures monotonously approach their pristine form as the separation 
between adjacent pores is increased. The densities of states (DOS) 
for the band structures presented are displayed in Figs.~\ref{azpore} (d, h). 

The energy band gaps in ZGNRs and AGNRs are known to arise from a staggered 
sublattice potential and a quantum confinement \cite{Son,Sols}, respectively, 
and they depend on the ribbon width and its functionalization 
\cite{Cervantes,Lu,Boukhvalov}.  The wave functions of the HOMO and LUMO 
bands in AGNRs, which contribute directly to the area near $E_f$, 
are localized at the center of the ribbons, keeping their edges chemically 
stable \cite{Cervantes}. In ZGNRs, these wavefunctions are localized at 
the ribbon edges. Consequently, when the SPs are positioned 
at the center of 11-AGNR a band gap opens in its band structure, while 
10-ZGNR remains metallic. The metallicity of ZGNRs is caused by 
flat bands present at $E_f$, originating from highly localized states 
formed at the zigzag edges \cite{Cervantes}, as discussed below.
When the pore is closer to the edge of 11-AGNR, its band structure is
influenced less, while no significant change is observed in 10-ZGNR.

\newpage
{\it Porous graphene:\ }

We continue to study PG that lacks explicit edges, in contrast to PGNRs. 
Graphene-based systems have a bipartite lattice \cite{Roche}, with 
$n_{A}$ and $n_{B}$ sites per unit cell in the $A$ and $B$ sublattices, 
leading to $n_{A}-n_{B}$ flat bands at the Fermi level \cite{Vanevic}. 
The metallicity caused by these bands is highly stable to any perturbation 
(see Figs.~\ref{azpore} (f, g)). Although, $n_{A}=n_{B}=1$ in graphene, 
introduction of pores or other defects may change its global sublattice 
balance, i.e. $n_{A}\neq n_{B}$. The relationship between edge-localized 
states, zigzag-like edges of GNR, and flat bands at the Fermi level was 
discussed 
in \cite{Nakada}. It turns out that even a small number of zigzag sites 
at the ribbon edge gives rise to highly localized states forming flat bands 
at $E_f$. These zigzag edges are characterized by a local sublattice 
imbalance and unbalanced $\pi$-electron spin density \cite{Ivanciuc}.
These localized states at zigzag rims were also studied in different types 
of pores, called ``anti-molecules", where a set of simple rules was shown
to link the net number of unpaired electrons with the degeneracy of flat 
bands \cite{Hatanaka,Furst,Vanevic,Palacios,Fazekas}. SPs, largely used 
in our study, keep the global sublattice balance of the unit cell, but
their short zigzag-like rims break the local balance and may still give 
rise to flat bands.

\begin{figure}[t]
\vspace*{0mm}
\hbox{\epsfxsize=85mm \epsffile{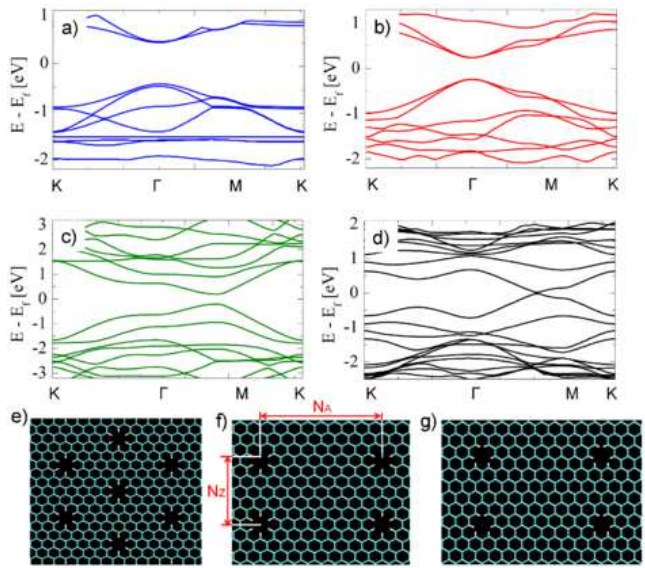}}
\vspace*{-3mm}
\caption{
\small (top) Band structure of a) a honeycomb SP-superlattice in e) and three
rectangular SP-superlattices with b) $N_{A}=15$, $N_{Z}=4$, c) $N_{A}=7$,
$N_{Z}=2$, and d) $N_{A}=7$, $N_{Z}=8$. (bottom) e) honeycomb 
SP-superlattice, f) rectangular SP-superlattice characterized by $N_{A}$,  
$N_{Z}$, g) rectangular superlattice with triangular-shape pores}
\label{Sgraphene}
\vspace*{-7mm}
\end{figure} 

In Figs.~\ref{Sgraphene} (a-d), we show the band structures of (edgeless) 
PGs with one honeycomb and three different rectangular SP-superlattices
defined by $N_{A}$, $N_{Z}$, and displayed in Figs.~\ref{Sgraphene} (e, f) 
\cite{Comment1}. 
Figure~\ref{Sgraphene} (g) also shows the studied rectangular superlattice 
with triangular-shape pores. The honeycomb SP-superlattice has flat bands 
around $E\approx -1.5$ eV (Fig.~\ref{Sgraphene} a), which are absent in the 
rectangular SP-superlattices (Figs.~\ref{Sgraphene} (b-d)). According to 
\cite{Vanevic}, such quasi-flat bands at non-zero energy might be ascribed to 
the local sublattice imbalance (globally $n_{A}=n_{B}$). However, 
the local sublattice imbalance due to SPs can not be the origin of flat 
bands in the band structure of the honeycomb SP-superlattice, since it 
has the same pores as the other structures with no flat bands. Therefore, 
the flat bands are most likely related with the honeycomb SP-arrangement, 
which might produce differently localized states. These results
show that the small number of zigzag sites at the very short SP-rim can not 
generate flat bands at the Fermi level or at its vicinity. In contrast, even 
the smallest triangular-shape pores break the global sublattice balance and 
generate an unbalanced $\pi$-electron density \cite{Ivanciuc}, associated with 
the appearance of rim-localized states and giving rise to flat bands at $E_f$ 
(not shown).

With these observations, we can now relate the band structures of PGNRs and 
PGs. Since SPs do not break the global sublattice balance in the PGNRs, their 
presence does not generate any new flat bands at $E_f$ (see Fig.~\ref{azpore});
on the contrary, we found that when AGNRs are perforated with triangular-shape 
pores, their band structure always contains flat bands at $E_f$. Even though 
the SPs do not create flat bands at $E_f$ in GNRs, they still influence their
band structure. While the band structures of AGNRs (Figs.~\ref{azpore} (b, c)) 
is influenced a lot, the band structures of ZGNRs with flat bands at $E_f$ 
(local sublattice imbalance caused by zigzag edges) can not be significantly 
modified by SPs. Therefore, we conclude that all porous ZGNRs are metallic 
(checked by calculations).

\begin{figure*}[t]
\vspace*{-3mm}
\hspace*{-2mm}
\hbox{\epsfxsize=150mm \epsffile{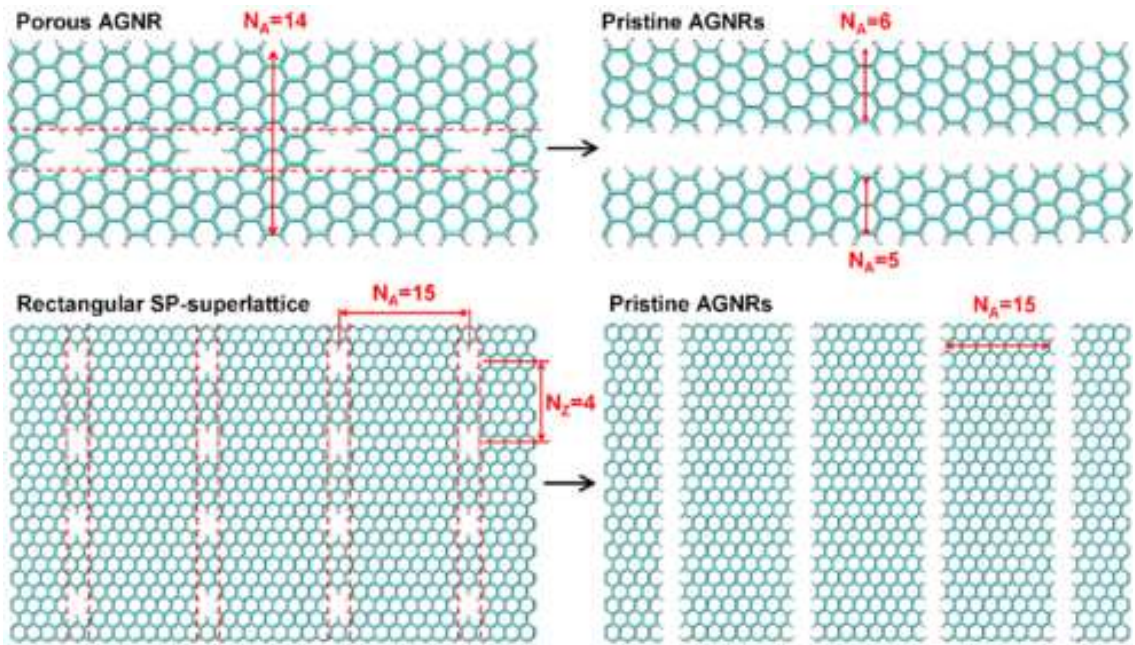}}
\caption{
\small
(top) Effective replacing of porous N-AGNR by ``daughter" pristine $N_{1}$- and
$N_{2}$-AGNR ($N=14$, $N_{1}=5$, $N_{2}=6$). (bottom) Effective replacing of 
rectangular SP-superlattice with $N_{A}>N_{Z}$ by set of prestine AGNRs 
($N_{A}=15$, $N_{Z}=4$).
}
\label{unzip}
\vspace*{-4mm}
\end{figure*}

We performed extensive {\it ab initio} calculations of electronic structures 
of PG-superlattices with different arrangements of the SPs. Interestingly, 
we found that rectangular PG-superlattices perforated with SPs and larger 
pores of honeycomb symmetry can be both metallic and semiconducting. It 
turns out that we can generalize these observations into a hypothesis that, 
in the first approximation, the electronic structure
of these superlattices has the same type of conductivity as many parallel 
AGNRs or ZGNRs (of effective widths $N_{A}$ or $N_{Z}$), depending on the 
ratio $\rho=N_{A}/N_{Z}$ (see Fig.~\ref{unzip} bottom). If $\rho \gg 1$, 
one can see the superlattice as being ``cut" into separated $N_{A}$-AGNRs, 
while for $\rho\ll1$ the same is true for separated $N_{Z}$-ZGNRs.

\begin{figure*}[t]
\vspace*{-1mm}
\hbox{\epsfxsize=180mm \epsffile{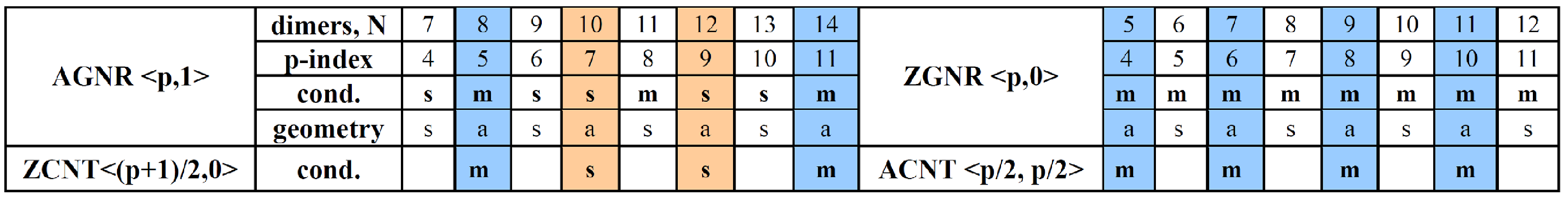}}
\vspace*{-3mm}
\caption{Correspondence rule for GNRs and CNTs. 
$N$ - number of dimers forming GNR, 
$p$ - index for the chirality vector in GNR and CNT;
cond. - conductivity property (semiconducting (s) or metallic (m)); geometry 
defines the symmetry with respect to the mirror plane perpendicular to the 
ribbon and containing its axis: symmetric (s), asymmetric (a). Examples of
AGNRs for $N = 7,...14$ and ZGNRs for $N = 5,...12$ are illustrated.} 
\label{Table}
\vspace*{-3mm}
\end{figure*}

The above hypothesis was largely confirmed by our follow up calculations.
For example, the conductivity in the PGs with $N_Z=2$ is the same as in 
the corresponding AGNRs: metallic for $N_A=5, 11, 17...$ and semiconducting
for $N_A=7, 9, 13, 15...$ (see also Fig.~\ref{Table}). If we continue
with the $N_A=9, 15, 21, ...$ semiconducting AGNRs and increase the initially 
small $N_Z=2$, we find that the PGs remain semiconducting for (roughly) 
$N_Z<N_A$, with the band gap shrinking with increasing $N_Z$, signaling the 
transition to the ZGNR-dominated metallic conductivity. If we continue with
the the $N_A=11$ metallic AGNRs, the PGs become metallic for all the $N_Z$,
since the ZGNRs that take over at $N_Z>N_A$ are all metallic. Finally, 
when we continue from $N_A=7, 13...$, the metallicity appears abruptly at 
$N_Z\ge 4$. In other cases, we expect that the transition between the AGNR and 
ZGNR-type of behavior occurs somewhere around the ``diagonal", $\rho=1$. 
Our calculations also show that the metal-semiconductor transitions 
predominantly occur in two regions of the Brillouin zones, as seen in 
Figs.~\ref{Sgraphene} (b-d), and the bands can be partially flat in 
the $k_y$ direction ($N_A=7, 13,..a.$).

{\it Porous nanotubes:\ } 

\begin{figure}[]
\hspace*{-2mm}
\hbox{\epsfxsize=87mm \epsffile{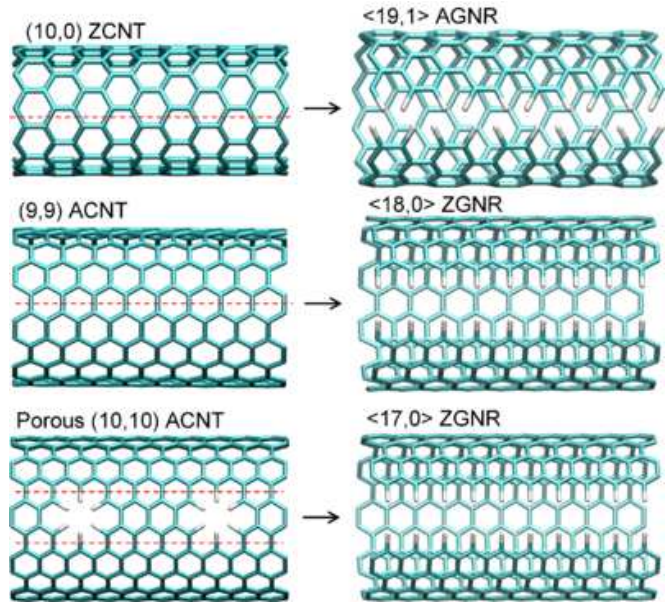}}
\caption{
\small
Cutting of: (top) (10,0)ZCNT into $\langle 19,1 \rangle$AGNR and (middle) 
(9,9)ACNT into $\langle 18,0 \rangle$ZGNR  (opening of the GNRs is 
schematically shown). (bottom) Removal of atoms from porous (10,10)ACNT 
leading to $\langle 17,0 \rangle$ZGNR.
}
\label{unzip1}
\vspace*{-6mm}
\end{figure}

It turns out that there is also a clear correspondence between pristine CNTs 
and structurally analogous GNRs (GNRs can be rolled into CNTs).
While ZGNRs indexed by $\langle p,0 \rangle$ are metallic 
for all $p$, only one third of AGNRs indexed by $\langle p,1 \rangle$ 
are metallic. Since, only one half of AGNRs can be rolled into ZCNTs, only 
one third ZCNTs are metallic \cite{Ezawa}. We found that this mapping can be 
generalized to the point that every metallic/semiconducting CNT corresponds
to a metallic/semiconducting GNR. The unambiguous relationship can be 
expressed in terms of the chirality index, $p$, and the number of GNRs 
carbon dimers, $N$, as follows
\begin{eqnarray}
\mbox{ZCNT}\langle p,0 \rangle \Rightarrow N\mbox{-AGNR},\ \ 
\mbox{where}\ \ N = 2p+2\, ,
\nonumber \\ 
\mbox{ACNT}\langle p,p \rangle \Rightarrow N\mbox{-ZGNR},\ \ 
\mbox{where}\ \ N = 2p+1\, .
\label{rule}
\end{eqnarray}
Intuitively, we can look at this CNT-GNR correspondence as a consequence 
of CNT ``cutting" (Figs.~\ref{unzip1} (top, middle)), which preserves the type 
of conductivity. The AGNRs and ZGNRs that do not match any CNTs are all 
semiconducting and metallic, respectively, as summarized in Fig.~\ref{Table}.

\begin{figure}[!b]
\vspace*{-6mm}
\hbox{\epsfxsize=87mm \epsffile{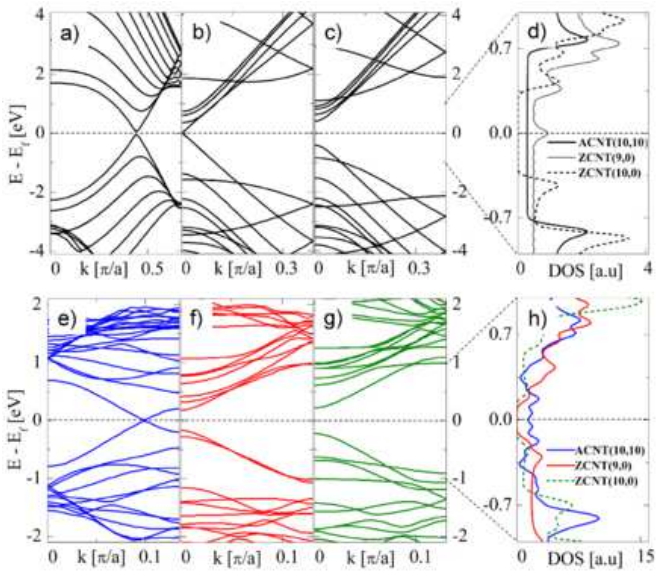}}
\caption{
\small
Band structures in pristine CNTs: 
a) ACNT (10,10), b) ZCNT (9,0), c) ZCNT (10,10); d) DOS.
Porous CNTs: e) ACNT (10,10), f) ZCNT (9,0), g) ZCNT (10,10); h) DOS.
The energy scales for (b,c) and (f, g) cases are the same.
}
\label{CNT}
\vspace*{-3mm}
\end{figure}

Analogously, porous CNTs might have band structures similar to porous GNRs. 
Tight-binding calculations predicted \cite{Sato} that a line of SPs 
(separated by $\sim12.8$ {\AA}) should cause band gap opening in ACNTs, 
whereas porous ZCNTs should be semiconducting regardless of the pore shape.  
These results are in contradiction with our {\it ab-initio} calculations, 
which show that the metallicity of pristine ACNTs 
(Fig.~\ref{CNT} a) is preserved in the porous ACNTs (Fig.~\ref{CNT} e), 
even for triangular-shape pores with clear zigzag-like rims. In metallic 
ZCNTs, the SP-perforation causes band gap opening, as shown in 
Figs.~\ref{CNT} (b, f), while in semiconducting ZCNTs, it causes band gap 
shrinking, as seen in Figs.~\ref{CNT} (c, g).

\newpage
\vspace{2mm}
{\bf \Large Discussion}

The above observations allow us to build a {\it unified model} that can predict 
the type of conductivity in porous nanocarbons perforated with SPs and other
pores that do not break the global sublattice balance. The model is based 
on the assumption that when NCs are perforated by a line of relatively 
close SPs, the type of conductivity in these PNCs is the same as in 
the (daughter) systems obtained from these NCs by removing all C atoms within 
a stripe going in the direction of the pores and having the same width as
the pores (all dangling C bonds are H-terminated). We call these modified NCs 
the daughter systems of the original NCs (two GNRs for PGNRs, one GNR for 
PCNTs, and a GNR-lattice for PG). 
This rule predicts that: (1) Porous ACNTs are metallic as the (daughter) 
ZGNRs; Fig.~\ref{unzip1} (bottom) shows effective replacing of porous ACNT 
by pristine ZGNR. Perforating the resulting ZGNRs (and the other half 
of ZGNRs that can not be rolled up into CNTs) gives two metallic ZGNRs, 
preserving the ZGNR-metallicity. (2) Porous ZCNTs may give semiconducting or 
metallic AGNRs. Cutting all the AGNRs may give pairs of AGNRs with any 
conductivity. These results were confirmed by {\it ab initio} calculations.

We now use these rules to predict metallicity in porous AGNRs with SPs.
We assume that their band gaps are $E_{BG} \simeq {\rm min}
\left(E_{1BG};\, E_{2BG}\right)$, where $E_{1BG}$, $E_{2BG}$ are 
band gaps of their two daughter AGNRs (see Fig.~\ref{unzip} (top)).
With this inference, we can derive an analytical expression describing 
the dependence of the band gap on the width of the porous AGNRs. 
For simplicity, we consider SPs positioned in the middle of AGNRs of 
the width of $W=a\,\sqrt{3}\,(N-1)/2$, where $a$ is the C-C distance 
and $N$ is the number of dimers. By evaluating the widths of the pristine 
daughter AGNRs, we find that porous $N$-AGNRs are potentially metallic 
if the number of C-C dimers is given by at least one of these equations
\begin{eqnarray}
N& =& \left(6k+11+(-1)^k\right)/2\, , \ \ N = 6k+3+2\,(-1)^k\, ,
\nonumber \\
N& = & 12k+9,\, \ \  (k=0, 1, 2...) \, ,
\label{cut}
\end{eqnarray}
i.e., if $N=5, 6, 7, 8, 9, 12, 14, 17, 18, 19, 20,...$. \\

\begin{figure}[!h]
\hbox{\epsfxsize=75mm \epsffile{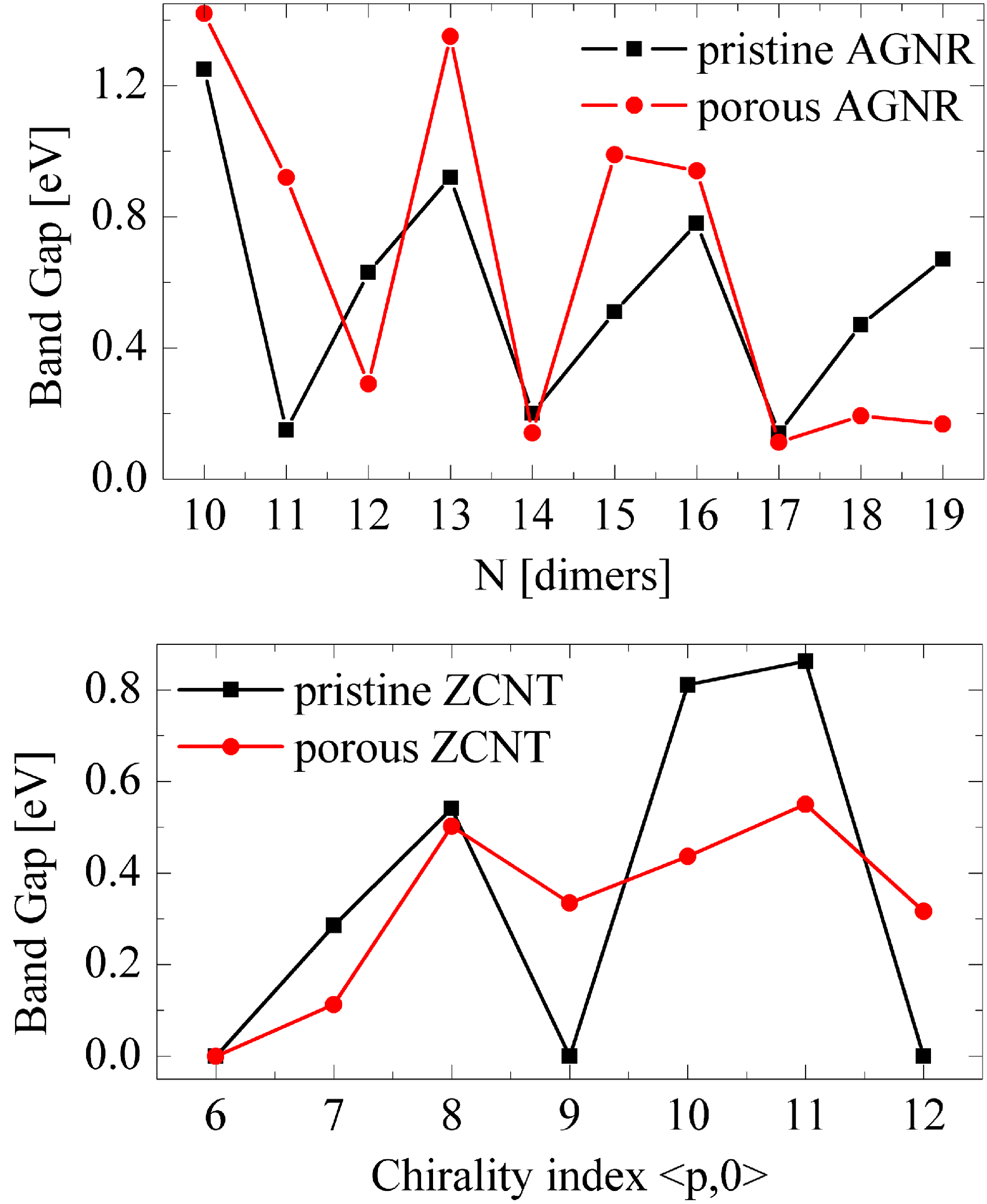}}
\caption{
\small
(top) Dependence of the band gap in the pristine and porous AGNR on the 
number of dimers (the central position of the SP). (bottom) The same
dependence in ZCNT on the chirality index.
}
\vspace*{-3mm}
\label{BG}
\vspace*{-0mm}
\end{figure}

In Fig.~\ref{BG} (top), we compare the {\it ab-initio} energy band gaps 
in pristine \cite{Barone,Yu} and porous AGNRs to validate the above 
model. In contrast to the pristine ribbons, where the metallic points 
emerge with the period of $3$, ($N_{met}=3m+2$), the band gaps of porous 
AGNRs have a more complex dependence. Nevertheless, the positions 
of the band gap minima agree with Eqns.~\ref{cut}. 

We can extend the assumptions used in Eqns.~\ref{cut} to PNCs perforated with 
larger and shifted honeycomb-like pores. Their presence may still be reduced to 
removing from the AGNRs a layer of atoms of the width given by the pore size,
where the minimum band gap of the two resulting AGNRs can determine the band gap
of the porous AGNR. For example, when the SP is shifted in the 11-AGNR by 
one honeycomb from the ribbon center, the two daughter pristine 4-AGNRs are 
replaced by 2-AGNR and 6-AGNR (all semiconducting). This should lead to
a band gap shrinkage, in agreement with our {\it ab-initio} calculations, 
presented in Figs.~\ref{azpore} (b, c). Alternatively, we can replace 
the SP by a double-size hexagonal pore with 24 C-atom excluded. If the 11-AGNR 
and 12-AGNR are perforated by such pores, they become semiconducting, since 
their cutting leads to semiconducting 2-AGNRs and 2- and 3-AGNR, respectively.
These results are in agreement with {\it ab-initio} calculations,
giving in 11-AGNR and 12-AGNR the band gaps of $1.1$ eV and $1.18$ eV, 
respectively. We have also tested the triple-size hexagonal pore (54
C-atoms excluded) in order to check how its long rims affect band
structure of GNRs. Our calculations show that no additional features
(e.g. flat bands at the Fermi level) appear when the GNR are perforated
by these pores. In contrast, when these AGNRs are perforated with SPs 
in the ribbon center, only the 11-AGNR is semiconducting.

Finally, we discuss porous ZCNTs that can have any conductivity. 
In Fig.~\ref{BG} (bottom), we present the energy band gap of porous 
ZCNTs in dependence on the chirality index, $p$. It exhibits similar 
periodicity as in the pristine ZCNTs. However, the model is not reliable 
in porous ZCNTs. For example, the porous ZCNT(7,0) and ZCNT(8,0) 
have band gaps similar to the daughter 11-AGNR and 13-AGNR, respectively. 
But the same is not true for the porous ZCNT(9,0) and ZCNT(10,0) paired 
with the daughter 15-AGNR and 17-AGNR, respectively. In principle,
this failure might be caused by the fact that the AGNRs are not 
calculated deformed as the corresponding daughter ZCNTs \cite{Zhang,Guo}. 
However, our calculations show that the bent 17-AGNR has almost the same 
band gap as the pristine 17-AGNR. Therefore, a more quantitative approach
needs to be used here.

\begin{figure}[!h]
\vspace*{-0mm}
\hspace*{-2mm}
\hbox{\epsfxsize=89mm \epsffile{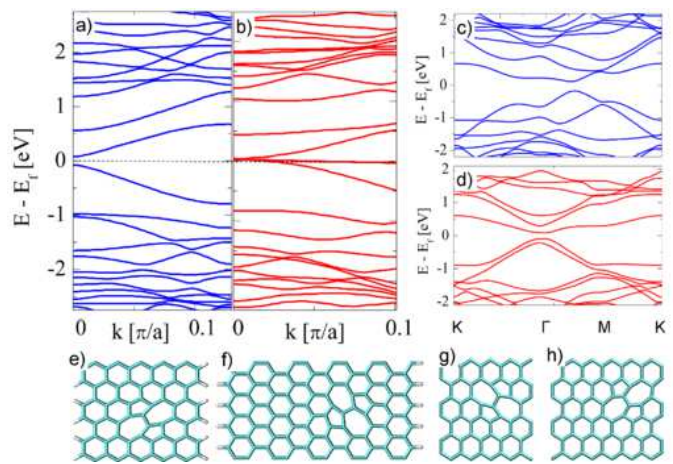}}
\caption{
\small
Band structures of nanocarbons with periodic 55-77 Stone-Wales defects:
a) 11-AGNR, b) 10-ZCNT, c) graphene superlattice with $N_A=7$ and $N_Z=4$,
d) graphene superlattice with $N_A=9$ and $N_Z=4$. Figures (e-h) show
unit cells for the respective cases.
}
\vspace*{-6mm}
\label{DNC}
\vspace*{3mm}
\end{figure}

It is of interest to see if other types of periodic modifications can also
be used to tune the band structures of nanocarbons. To briefly examine this
idea, we have replaced SPs by Stone-Wales 55-77 defects \cite{Banhart}. In 
Fig.~\ref{DNC}, we show the band structures of 11-AGNR, 10-ZGNR, and 
graphene superlattices modified in this way. The periodic array of SW 
55-77 defect in 11-AGNR leads to small bang gap opening, as shown 
Fig.~\ref{DNC} (a), in analogy to 11-AGNR with SPs (Fig.\ref{azpore} b). 
The band structure of 10-ZGNR, shown in Fig.~\ref{DNC} (b), is not sensitive 
to this perturbation, as in the SP-perturbations (Fig.\ref{azpore} f). 
On the other hand, when we replace in graphene superlattices SPs with the 
SW 55-77 defects, we can obtain qualitatively different band structures.
In particular, the band structure of graphene modified by SW 55-77 defects
in the array with $N_A=7$ and $N_Z=4$ (Fig.~\ref{DNC} c) is similar to that 
of the SP-superlattice with the same $N_A$ and $N_Z$, but here we also observe 
opening of a small band gap. In Fig.~\ref{DNC} (d), we show the 
band structure of the SW-graphene superlattice ($N_A=9$ and $N_Z=4$), 
which is semiconducting as the PG-superlattice with the same $N_A$ and $N_Z$. 
These observations show that periodic defects could also be used to tune 
band gaps in nanocarbons, but the rules might be slightly different.

For completeness, we have recalculated some of the above structures
including spin polarization (we used a set of LDA and GGA
functionals). It turns out, the band structures of NCs can be modified
by the spin polarization (zigzag edges) \cite{Palacios}, but the presence of SPs
does not introduce additional magnetic features beyond the changes
described already in the non-magnetic calculations. Interestingly, the
presence of arrays of SW (55-77) defects in ZGNRs can alternate
mutual orientation of the magnetic moments localized at the opposite
edges, due to topological changes in the sublattices.

In summary, we have developed a unified picture of band structures in PNCs. 
Although, the proposed approach successfully describes band structures 
in many PNCs, it could be further refined to account for quantization, 
spin degrees of freedom, pore type, and chirality. Similar observations were 
made in nanocarbons perturbed by periodic SW defects. Precise knowledge of 
electronic structures of these materials is essential for their 
applications in electronics, optics, molecular sensing, and other fields.

\vspace{2mm}
{\bf \Large Methods}

\small
We study the PNCs {\it ab initio}, using SIESTA 3.0-beta-15 \cite{Sanchez} 
in supercells of ($>40$ atoms), and neglect spin degrees of freedom. The length
of unit cells for porous 11-AGNR and 10-ZGNR is $12.78$ {\AA} and $12.3$ {\AA},
respectively. The size of the supercells of PG used in our calculations varies 
between $12.3 \times 8.52$ {\AA} (40 atoms) and $22.13 \times 21.3$ {\AA} 
(180 atoms).  We use the Perdew-Zunger LDA functional \cite{Perdew} and 
pseudopotentials with the cutoff energy of 400 Ry. The calculations are done 
within the eigenvalue tolerance of $10^{-4}$ eV, using the DZP basis 
(double-zeta basis and polarization orbitals, 13 and 5 orbitals for C-atom and 
H-atom, respectively). The Brillouin zones of the unit cells are sampled by 
the Mankhorst-Pack grid \cite{Monkhorst} with the spacing between the 
$k$-points of $\Delta k <0.01$ {\AA}$^{-1}$. Geometry optimization is carried 
out for all the PNCs within the conjugated gradient algorithm, until all the 
forces are $F<0.04$ eV/{\AA} and the stress in the periodic direction is 
$\sigma < 0.01$ GPa. We use a.u. for $k$ in the band structures, where 
only one half of the Brillouin zone is shown (symmetry). 

\noindent 


\vspace{3mm}
\noindent 
{$^*$\small The address for correspondence: pkral@uic.edu}

\vspace{5mm}
{\bf Acknowledgments}\\
We acknowledge support from the NSF CBET-0932818 grant. The calculations
were performed at the NERSC, NCSA, and CNM supercomputers. We thank
Niladri Patra, Irena Yzeiri, and Cathy Skontos for technical help.

\vspace{5mm}
{\bf Authors Contributions}\\
A.B. and P.K. wrote the main manuscript text. A.B. performed all 
the calculations. All authors discussed the results and commented 
on the manuscript.

\vspace{5mm}
{\bf Additional Information}\\
The authors declare no competing financial interests.

\clearpage


\begin{thebibliography}{99}
\bibitem{Comment1} We use rectangular unit cells, and 
calculate the band structure along the direction $K\to{\Gamma}\to 
M\to K$ in the reciprocal space, where $M=(k_x=\pi/a,\, k_y=0)$.
\end{thebibliography}
\end{document}